\documentclass[useAMS]{mn2e}

\newif\ifAMStwofonts

\usepackage{graphicx}

\title[X-ray polarisation in ``changing look'' AGN -- The case study of NGC~1365]
      {Modelling the X-ray polarimetric signatures of complex geometry: the case study of the ``changing look'' AGN NGC~1365}\author[F. Marin et al.]
      {F.~Marin$^1$\thanks{E-mail: frederic.marin@astro.unistra.fr},
       D. Porquet$^1$, R. W.~Goosmann$^1$, M.~Dov{\v c}iak$^2$, F. Muleri$^3$,
       N. Grosso$^1$
       \newauthor 
       and V. Karas$^2$ \\
       $^1$Observatoire Astronomique de Strasbourg, Universit{\'e} de Strasbourg, 
       CNRS, UMR 7550, 11 rue de l'Universit{\'e}, 67000 Strasbourg, France \\
       $^2$Astronomical Institute of the Academy of Sciences, Bo{\v c}ni
       II 1401, 14131 Prague, Czech Republic \\
       $^3$INAF/IAPS, Via del Fosso del Cavaliere 100, I-00133 Roma, Italy}
\date{Accepted 2013 September 04. Received 2013 August 30; in original form 2013 May 16}

\pagerange{\pageref{firstpage}--\pageref{lastpage}}
\pubyear{2013}

\begin{document}

\maketitle

\label{firstpage}

\begin{abstract}

``Changing look'' Active Galactic Nuclei (AGN) are a subset of Seyfert galaxies characterized by
rapid transitions between Compton-thin and Compton-thick regimes. In their Compton-thin state, the central 
engine is less obscured, hence spectroscopy or timing observations can probe their innermost structures. 
However, it is not clear if the observed emission features and the Compton hump are associated with 
relativistic reflection onto the accretion disc, or complex absorption by distant, absorbing gas clouds 
passing by the observer's line-of-sight. Here, we investigate these two scenarios under the scope
of X-ray polarimetry, providing the first polarisation predictions for an archetypal ``changing look'' 
AGN: NGC~1365. We explore the resulting polarisation emerging from lamp-post emission and scattering off 
an accretion disc in the immediate vicinity of a supermassive black hole. The computed polarisation 
signatures are compared to the results of an absorption-dominated model, where high column density gas partially 
covers the central source. While the shape of the polarisation spectrum is similar, the two models differ in net 
polarisation percentage, with the relativistic reflection scenario producing significantly stronger polarisation. 
Additionally, the variation of the polarisation position angle is distinctly different between both scenarios: the 
reflection-dominated model produces smooth rotations of the polarisation angle with photon energy whereas circumnuclear 
absorption causes an orthogonal switch of the polarisation angle between the soft and the hard X-ray bands. 
By comparing the predicted polarisation of NGC~1365 to the detectability levels of X-ray polarimetry 
mission concepts proposed in the past,  we demonstrate that with a large, soft X-ray observatory or a medium-sized mission 
equipped with a hard (6 -- 35~keV) polarimeter, the correct interpretation would be unambiguous.
\end{abstract}


\section{Introduction}

AGN are complex objects, hosting several media over a broad range of location. Radiation from the central,
ionising source is absorbed and reprocessed by a variety of scattering regions such as the accretion disc, 
the Broad Line Region (BLR), the Warm Absorber (WA), the molecular torus, and the Narrow Line Region (NLR), 
imprinting the resulting X-ray spectrum of AGN. For example, the type~1.8 AGN NGC~1365 (z~=~0.00547), 
also known as the Great Barred Spiral Galaxy, is an intriguing source displaying the most
dramatic X-ray spectral changes observed so far in an AGN: the source switched from reflection-dominated 
(Compton-thick) to transmission-dominated (Compton-thin) and back in just a few days to week time-scales 
(Risaliti et al. 2005). NGC~1365 can be classified as one of the most extreme examples of ``changing look``
AGN. This has been attributed to variation in the line-of-sight of cold absorber rather than to extreme
intrinsic emission variability. Such rapid time-scales of these X-ray eclipses suggest that these 
absorbers (N$_{\rm H}\sim$ 10$^{23}$ -- 10$^{24}$\,cm$^{-2}$) are located on compact scales consistent within 
the inner BLR and the outer part of the accretion disc. Such behaviour is now observed in numerous other 
sources, e.g. NGC ~4388 (Elvis et al. 2004), NGC~7674 (Bianchi et al. 2005), NGC~4151 (Puccetti et al. 2007), 
NGC~7582 (Bianchi et al. 2009), UGC~4203 (Guainazzi et al. 2002; Risaliti et al. 2010), NGC~4051 
(Guainazzi et al. 1998; Lobban et al. 2011), 1H~0419-577 (Pounds et al. 2004), NGC~454 
(Marchese et al. 2012), IRAS~09104+4109 (Piconcelli et al. 2007; Chiang et al. 2013), H0557-385 
(Longinotti et al. 2009), and the Phoenix galaxy (Matt et al. 2009).
Besides, it  has been shown that NGC~1365 exhibits an apparent broad Fe~K$\alpha$ fluorescent emission line 
(Brenneman et al. 2013; Risaliti et al. 2013). Two alternative origins are proposed to explain the 
shape and variability of this feature, general relativistic effects (Risaliti et al. 2013) or partial-covering 
absorptions (Miller \& Turner 2013). In the former scenario, reflection of hard X-rays off the inner disc, 
reaching down to the innermost stable circular orbit (ISCO), produces fluorescent emission, subsequently broadened 
by general relativistic and Doppler effects. Integrating the reprocessed emission across the disc
constitutes the blurred line that is sensitive to the spin of the SMBH (Fabian et al. 1989, Laor 1991, 
Dov{\v c}iak et al. 2004, Brenneman \& Reynolds 2006, Dauser et al. 2010). In the latter scenario, complex 
absorption due to clouds of absorbing gas, partially covering the central AGN source, is responsible for
both the flux variability and the apparent broadening of the Fe line (Inoue \& Matsumoto 2003, Miller, 
Turner \& Reeves 2008, 2009). Continuum radiation transmitted and scattered through circumnuclear 
medium carves out the extended red wing, which leads to systematically lower measurements of the SMBH spin 
than obtained by the relativistic reflection method. Interestingly, the variability behaviour of the Fe K 
line is contrary to that of the relativistic line seen in the type-1 Seyfert galaxy MCG-6-30-15 
(Fabian et al. 2002), for which light bending effect has been proposed (Fabian \& Vaughan 2003; 
Miniutti et al. 2003). However, this can alternatively be explained by an increase of the column density 
of the cold absorber from 2008 to 2010 that would reduce the red wind of the line as pinpointed by 
Brenneman et al. (2013).

In this context, we aim to illustrate the potential of X-ray polarimetry observations as a powerful tool 
to test the complex geometry in ``Changing look'' AGN, using NGC~1365 as a study case.

\section{Polarisation induced by relativistic reflection}
\label{sec:reflect}

\begin{figure*}
 \centering
 \includegraphics[trim =10mm 200mm 20mm 12mm, clip, width=16cm]{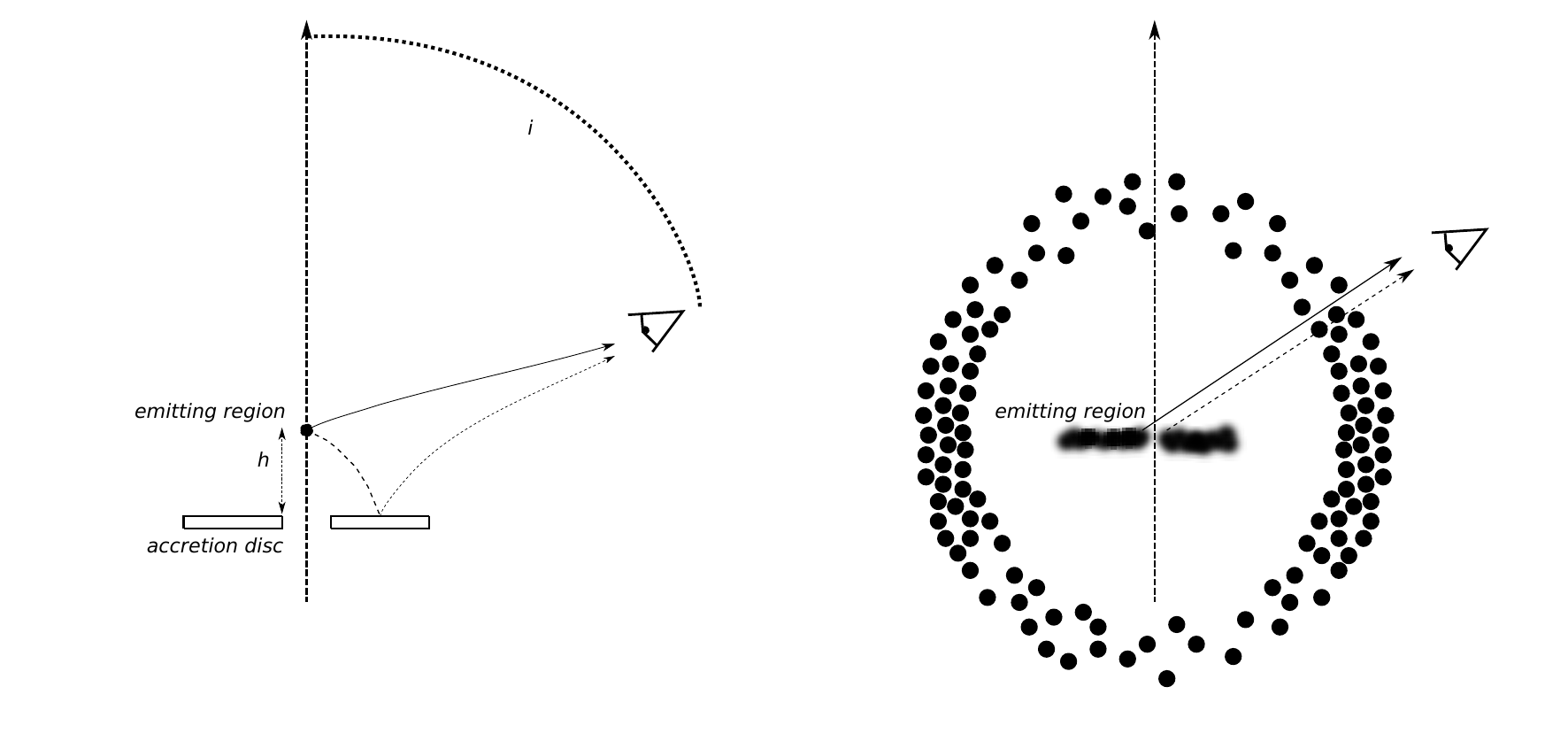}
 \caption{Schematic view of the two scenarios considered. $Left$:
   reflection with a lamp-post geometry and light-bending. $Right$:
   partial covering with circumnuclear absorbers.}
 \label{Models}%
\end{figure*}

The combined \textit{XMM-Newton} / \textit{NuSTAR} observation of NGC~1365 
(Risaliti et al. 2013) puts forward a scenario in which the Fe~K$\alpha$ emission line originates from a 
cold accretion disc surrounding a SMBH of mass $M = 2\times10^6$~M$_\odot$, associated with 
a dimensionless spin parameter $a = 1$ (extreme Kerr BH). Their model assumes that the X-ray source, 
illuminating the accretion disc, is localised on the rotation axis, at 2.5~gravitational 
radii above the black hole. The broadening of the Fe~K$\alpha$ fluorescence line is then a 
consequence of rapid motion and general relativistic effects close to the SMBH. Using the 
method already described in details in Dov{\v c}iak et al. (2011) and used in Marin et al. (2012a) 
to model X-ray polarisation spectra emerging from the Seyfert-1 MCG-6-30-15, we now apply it to 
the case of ``changing look'' AGN in their Compton-thin regime. 

Relying on the prescribed reflection model (Risaliti et al. 2013), we run Monte Carlo radiative 
transfer simulations of the 1 -- 100~keV polarisation emerging from NGC~1365. The input spectrum 
($>$ 10$^9$ photons) has a power-law shape $F_{\rm *}~\propto~\nu^{-\alpha}$ with a spectral 
index $\alpha$ = 1.2 (Risaliti et al. 2009, Risaliti et al. 2013). 
The intrinsic flux is unpolarised and emitted isotropically. The re-emitted intensity as a 
function of incident and re-emission angle is computed by the Monte-Carlo radiative transfer 
code {\it NOAR} (Dumont et al. 2000), the local polarisation being estimated according to the 
transfer equations of Chandrasekhar (1960). The local, polarised reflection spectra are then 
combined with the {\it KY}-code (Dov{\v c}iak et al. 2004), that conducts relativistic ray-tracing 
between the elevated source, the disc, and the distant observer (see Fig.~\ref{Models}, left-hand panel).

The resulting polarisation signatures, as a function of photon energy, are plotted in 
Fig.~\ref{Results} (red dashed line). $P$ is the degree of linear polarisation emerging from the models 
and $\Delta\Psi$ the rotation of the polarisation position angle with respect to a convenient 
average of the polarisation position angle over the depicted energy band. An "absolute orientation" 
of the polarisation plane is difficult to measure, and we thus focus on the variation of 
$\Delta\Psi$ with photon energy.

Considering an inclination of $i$ = 60$^\circ$ for NGC~1365 (Risaliti et al. 2013), 
the reflection model is found to produce relatively high degrees of polarisation 
(P~$\ge$~1~\%) in the 3 -- 100~keV band, with a maximum $P$ in the Compton hump,
where multiple scattering dominates. Below 10~keV, $P$ decreases sharply, unlike an energy-independent 
Newtonian case, as the local disc polarisation is affected by the relativistic energy shift 
of the photons and impacts the resulting polarisation seen by a distant observer 
(Dov{\v c}iak et al. 2004). The energy-dependence of $P$ in the soft X-ray band is accentuated 
by a dilution mechanism caused by the source, set to favour the production of soft X-ray photons: 
even if strong gravity effects tend to bend photons back to the disc, a fraction of
the input, unpolarised radiation merges with the (reprocessed) polarised photons and dilutes the $P$ signal.
Thus, the minimum polarisation degree produced by disc reflection is $\ge 0.1$~\% above 1~keV, 
$\ge 1$~\% between 3 and 10~kev, and up to 10~\% in the Compton hump regime.

Interestingly (Fig.~\ref{Results}, bottom, red dashed line), the relativistic reflection scenario 
exhibits an energy-dependent $\Delta\Psi$ that varies significantly across the Fe K$\alpha$ line, a 
behaviour already expected from X-ray polarisation simulations in the type-1 AGN MCG-6-30-15 
(Marin et al. 2012a). The feature is related to the energy-dependent albedo and scattering phase 
function of the disc material. $\Delta\Psi$ presents a smooth and continuous variation by $\sim 5^\circ$ 
around the iron line and increases monotonically upward 20~keV.

\section{Pure absorption and Compton scattering by a cloudy medium}
\label{sec:abso}

\begin{figure}
 \centering
 \includegraphics[trim = 10mm 10mm 42mm 30mm, clip, width=8.5cm]{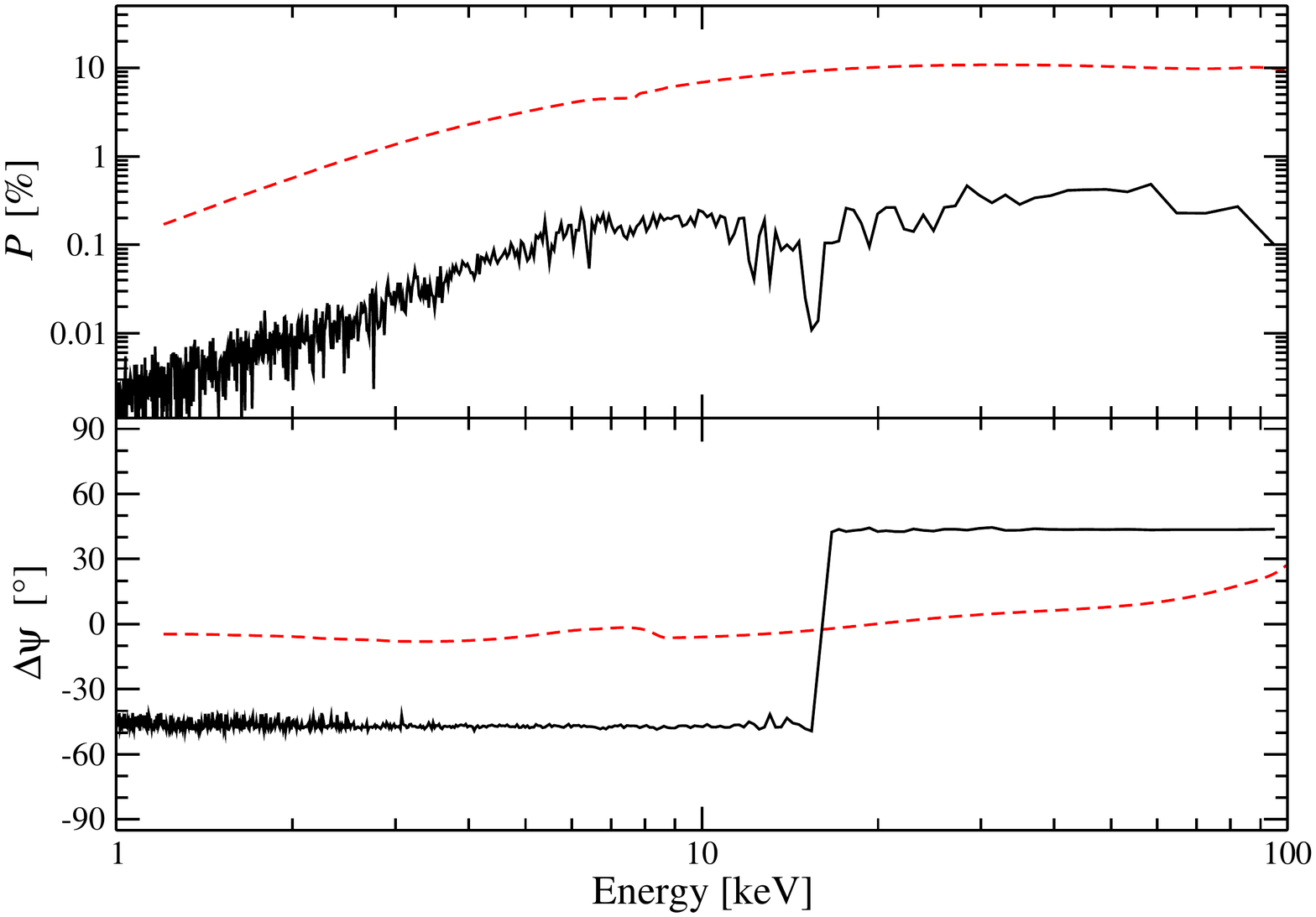}
 \caption{Percentage of polarisation $P$ and variation of the
   polarisation angle $\Delta\psi$ with respect to its mean as a
   function of the energy. $Legend$: a fragmented absorption region (solid line) 
   and a relativistic reflection model with an extreme Kerr SMBH with $a=1$ 
   (red dashed line). Both models are viewed at an inclination of 
   $i$ = $60^\circ$ to the axis of symmetry. }
 \label{Results}%
\end{figure}

Taking into account that the changing nature of NGC~1365 is explained by the presence of a 
clumpy distribution of gas around the central engine, we now investigate the resulting polarisation 
signal arising from an absorption-dominated model. We rely on the parametrisation of Miller \& Turner (2013) 
to produce a clumpy environment favourable to X-ray eclipses. Our goal is not to prove that the model of
Miller \& Turner (2013) provides the same spectral energy distribution (SED) as the one proposed 
by Risaliti et al. (2013), reproducing both the flux variability and the apparent broadening of the Fe line. 
We rather examine the polarisation induced by a cloudlet distribution that is supposed to be competitive 
with its relativistic counterpart. 

Using the latest version of {\sc stokes} (Goosmann \& Gaskell 2007; Marin et al. 2012b), 
we computed the resulting polarisation of a cloud distribution such as the ones presented 
in Karas et al. (2000) and Miller \& Turner (2013). The model set-up is summarised in
Fig.~\ref{Models} (right-hand panel). The central, X-ray source is defined as a geometrically thin, 
emitting slab that may represent the so-called hot inner flow. The emitting region is insensitive to
scattering and does not reach down to the ISCO in order to avoid a possible confusion with 
relativistic disc reflection. Inside a spherical shell constrained between $r_1 = 10$ and $r_2 = 20$ (in
units of spheres radius) from the model's origin, a random distribution of 1 000 scale-independent, 
Compton-thick, absorbing spheres with equal radius and constant density, was generated. The covering 
factor of the cloud distribution varies systematically from the poles towards the equator, as suggested 
by the unified scheme of AGN, and is $\sim 0.55$ for $i = 0^\circ$, and $> 0.75$ for $i > 60^\circ$
(Antonucci 1993; Urry \& Padovani 1995). The column density of the absorbing, cold medium in a single cloud is equal 
to 1.5$\times10^{24}$~cm$^{-2}$, a value chosen by to maximise the observed hard excess. 
For coherency, we restrain ourselves to the same input power-law spectrum as in Sect.~\ref{sec:reflect}.
This parametrisation is in perfect agreement with the modelling of Miller \& Turner (2013), allowing us, 
in the mean time, to examine the net polarisation of ``changing look'' AGN and explore the polarimetric 
signatures of their theoretical, absorption-dominated model.

We find that, in the case of an absorption-dominated model (Fig.~\ref{Results}, top), the
resulting polarisation degree is energy-dependent (solid black line), with a maximum 
$P$ observed in the Compton hump ($P \le$ 0.5~\%). Close to 15~keV, the local polarisation 
strongly diminishes due to a flip of the photon position angle, as perpendicular and parallel photon's 
position angles cancel each other (see next paragraph). Finally, $P$ gradually decreases below the iron 
fluorescence line energy, as the dilution trend is strengthened by the absence of strong gravity effects
bending photons back to the disc. As forward scattering dominates, we naturally find that the polarisation 
produced by a clumpy distribution of cold matter is about 15 times weaker than the relativistic model.

\begin{figure}
 \centering
 \includegraphics[trim = 105mm 210mm 25mm 22mm, clip, width=8cm]{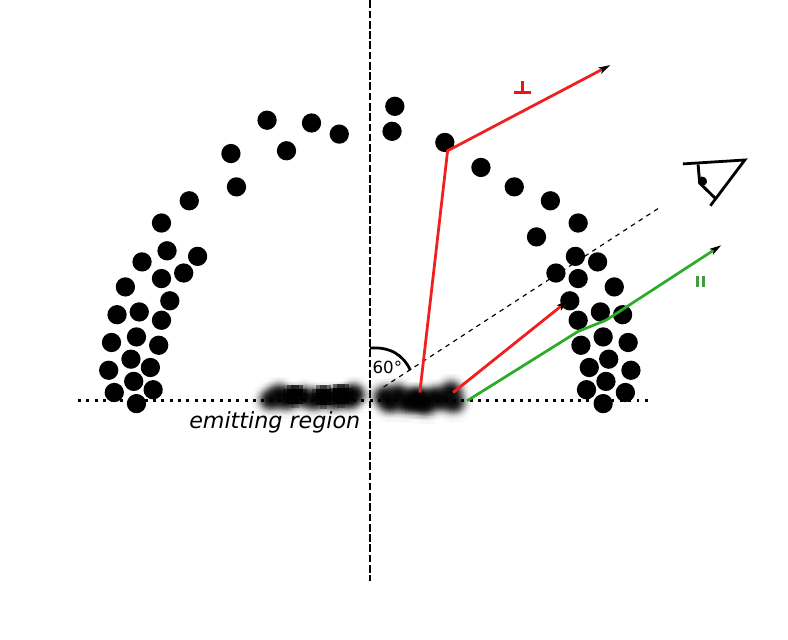}
 \caption{Schematic illustration of the dominant polarisation mechanism
	  for absorption and Compton scattering by a cloudy medium. Soft X-rays 
	  (red lines) are absorbed along the highly covered equatorial plane; to 
	  escape, radiation reprocesses on polar clouds and the average direction 
	  of the electric field vector of the scattered radiation is perpendicular 
	  to the axis of the system. Hard X-rays (green line) face a lower electron 
	  scattering cross-section and easily escape the circumnuclear medium by 
	  multiple scattering, aligning their electric field vector to the projected
	  axis of the system. Thus, a $\Psi$ rotation between the two regimes occurs 
	  at a model-dependent energy.}
 \label{Scattering}%
\end{figure}

The most interesting results come from the variation of the polarisation position angle (Fig.~\ref{Results}, bottom). 
The cloud distribution examined in this Letter (solid black line) shows an instantaneous $\Delta\Psi$ transition 
by 90$^\circ$ from the soft (below 15~keV) to the hard X-ray band, associated with a local diminution of $P$. In the 1 -- 15~keV
band, radiation is not energetic enough to scatter through the highly covered equatorial plane and becomes absorbed. 
Photons escaping from the system scatter on polar clouds, with their normal to the scattering plane perpendicular 
to the axis of the system, resulting in a polarisation position angle $\Psi$ = 0$^\circ$ (Marin \& Goosmann 2011; Marin et al. 2012b). 
Above 15~keV, the Compton scattering phase function favours scattering towards the forward direction and, at the same 
time, the albedo is strong so that photons escape along the observer’s viewing angle without suffering from heavy 
absorption. The planes of each scattering are predominantly parallel to the equatorial plane of the system, aligning their 
normal to the projected axis of the system. Thus, the polarisation vectors of the hard X-rays are orthogonal from those of soft X-rays, 
on average, and the net polarisation position angle $\Psi$ is equal to 90$^\circ$ (see Fig.~\ref{Scattering}). The 
competition between scattering in polar and scattering in equatorial clouds, leading to parallel or perpendicular 
polarisation angle, is energy-dependent and results in an orthogonal switch specific to the prescription. The energy of 
the $\Psi$ rotation is a complex function of opacity and covering fraction, and will be different for other parametrisations.

\section{Observational prospects}
\label{Prospects}

We showed, in a previous publication centred around MCG-6-30-15 (Marin et al. 2012b), that a small 
pathfinder mission could be able to distinguish between relativistic reflection and transmission 
through absorbing media. But, is this still true for NGC~1365, a rather different object with a 
much higher (60$^\circ$) inclination, a variable behaviour of its iron line red-wing, and extreme 
Compton-thick to Compton-thin regime transitions? Taking into account an average flux in the X-ray
band ($\sim$ 1.0~mCrab in the 2 -- 10~keV band, Brenneman et al. 2013, and $\sim$ 2.62 $\pm$ 0.50
mCrab in the 17 -- 60~keV band, Krivonos et al. 2007), we extend the sample of already proposed X-ray 
mission concepts of different sizes which included a polarimeter to investigate if any polarisation 
signal could be detected\footnote{We stress that our results are to be intended as a first estimate 
of the sensitivity which could be reached by state-of-the-art instruments in some exemplary mission 
profiles. The actual observability of the effects discussed in this paper, especially in case of 
the absorption scenario for which the expected polarisation signal is very low, will depend on a 
number of instrumental characteristics, i.e. systematic uncertainties, observation time, energy range, 
etc. It is difficult to take account of them at this stage because currently there are no X-ray 
polarimeters approved for the launch; notwithstanding, our results give an idea of what is 
reasonable to achieve in different mission profiles.}:

\begin{itemize}

  \item {\it XIPE}, the S-class X-ray Imaging Polarimetry Explorer (Soffitta et al. accepted), designed to be loaded
	with two 2 -- 10~keV Gas Pixel Detectors (GPD, Costa et al. 2001, Bellazzini et al. 2006, Bellazzini \& Muleri 2010).

  \item The M-class New Hard X-ray Mission {\it NHXM} (Tagliaferri et al. 2012a,b). One of the four telescopes aboard would 
	be coupled with two GPD instruments: a low energy polarimeter (LEP, 2 --10~keV) and a medium energy polarimeter (MEP,
	6 -- 35~keV), for a total coverage of 2 -- 35~keV.

  \item The L-class International X-ray Observatory {\it IXO} (Barcons et al. 2011) included, 
	among a few other instruments, the 2 -- 10~keV X-ray polarimeter XPOL (Bellazzini et al. 2006). 
\end{itemize}

\begin{figure}
 \centering
 \includegraphics[trim = 10mm 10mm 42mm 30mm, clip, width=8.5cm]{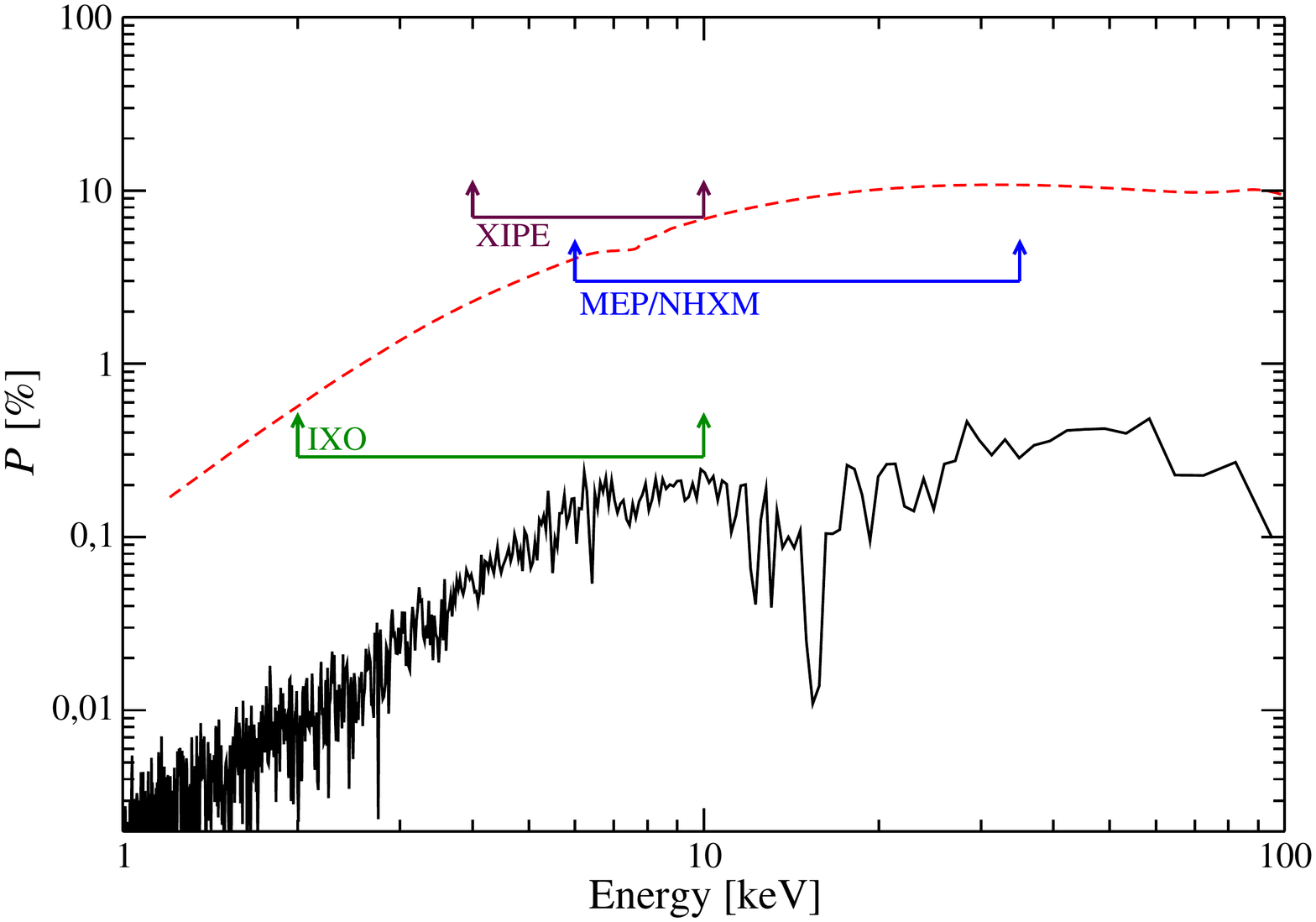}
 \caption{Representative minimum detectable polarisation (MPD) of the two scenarios for a
   1~Ms observation of NGC~1365. The observer's line-of-sight lies at
   $60^\circ$ with respect to the symmetry axis. The MDP for 
   {\it XIPE} is represented by the maroon line, for {\it NHXM} in blue
   and for {\it IXO} in green. $Legend$: a
   fragmented absorption region (solid line) and a relativistic
   reflection model with an extreme Kerr SMBH with $a=1$ (red dashed
   line).}
 \label{Predictions}%
\end{figure}

In Fig.~\ref{Predictions}, we calculated the representative minimum detectable polarisation (MDP, whose formula 
can be found in Marin et al. 2012a) on the whole energy range of the instrument at 99~\% confidence level that the mission 
concepts would have reached pointing at NGC~1365. The observation time is estimated to 1~Ms, taking into account that the 
background flux is negligible with respect to the source flux. We assumed a Crab-like spectrum, that is a power law with an 
index of 2.05 and fluxes reported above. The rate of the source is calculated on a fine energy grid (0.1~keV) by 
multiplying the telescope collecting area, the detector efficiency and the source spectrum. Then, we multiply (for each energy) 
the square root of the rate by the modulation factor, adding the background (which is negligible by orders of magnitude in this 
case) and smearing the product for the energy resolution of the instrument. Eventually, we sum up the values of the product 
in the energy range of interest to derive the MDP for the selected interval. The MDP evaluated over representative energy 
bands are: $\sim$~7~\% (4 -- 10~keV) for {\it XIPE}, $\sim$~3~\% for {\it MEP/NHXM} (6 -- 35~kev) and $\sim$~0.3~\% 
for {\it IXO} (2 -- 10~keV).

We find that the predicted polarisation signal induced by relativistic reflection is below the polarisation detectability 
of {\it XIPE} (maroon line) so, for an exposure time of 1~Ms, a small, pathfinder mission such as {\it XIPE} could not measure 
the polarisation unless the observed polarisation is significantly (a factor of 2) higher than that predicted by the model. 
However, any detection would strongly support the relativistic model that produces a conservative polarisation degree (see 
Sect.~\ref{Conclusions}. Further indications could be deduced from the variation of the polarisation position angle around 
the iron line if the observed polarisation signal is significant enough. The blue line, representing the 
MDP of {\it MEP/NHXM}, indicates that the polarisation signal originating from the relativistic reflection case 
can be detected across the whole 6 -- 35~kev band. It covers in particular the iron line domain, where the $\Delta\Psi$ 
signature of strong gravity would be detected. Finally, for longer observing time or higher flux levels, {\it IXO} 
(green line) would have been the only mission to potentially detect the low polarisation originating from a 
``changing look'' AGN in the complex absorption scenario.

\section{Discussion and conclusions}
\label{Conclusions}

Due to its transient nature, the distribution of gas clouds around the central region of NGC~1365 
may vary, both in its covering factor and column density. Lowering the covering factor would result in a 
decrement of the net polarisation as dilution from the unscattered input spectrum would be enhanced. A 
different density parametrisation would also affect the polarisation degree: optically thinner reprocessing 
clouds would reduce the probabilities of scattering, thus decreasing $P$, while increasing the gas density would 
slightly increase the polarised flux until the circumnuclear matter becomes optically too thick for the radiation 
to escape. Nonetheless, since absorption-dominated models are based upon continuum transmission through gas, 
the overall polarisation degree is expected to be weaker than for (perpendicular) scattering off an accretion disc. 
One must note that the energy at which the $\Psi$ rotation occurs is also governed by the gas density, covering 
factor, and possibly inclination. We intend to investigate different realisations of a variety of ``changing look'' 
AGN in future work in order to explore the complex inter-combination of parameters that may lead to similar 
$\Psi$ rotation. This feature, if observable, could be used as a major constraint to probe the parameter space
of the cold gas. 

The relativistic reflection model employed to simulate the polarisation response of NGC~1365 is basic, yet instructive.
In a forthcoming publication, we intend to improve our relativistic models by taking 
into account radially structured surface ionisation and intrinsically polarised, non-axisymmetric irradiation. 
Both implementations are expected to increase the net polarisation, similarly to coronal emission that scatters off 
an ionised disc and disc return radiation over the potential well (Schnittman \& Krolik 2010; 2013).
Therefore, the modelling presented in this Letter produces a conservative, lower limit on the polarisation degree. 

The polarisation position angle across the iron line turns out to be a particularly strong indicator 
for the correct scenario -- if it can be constrained with a sufficiently powerful X-ray polarimetry mission
and/or long enough observations. The bottom panel of Fig.~\ref{Results} shows that in the case of 
relativistic reflection $\Delta\Psi$ follows a characteristic ``S-shape'' between 2 and 9~keV,
whereas in the absorption scenario the distribution of $\Delta\Psi$ is only bipolar. The detection of a 
few degrees swing could be within the reach of a large mission, for which the MDP is significantly lower 
than the expected degree of polarisation. The fact that in the absorption case only ``parallel'' or 
``perpendicular'' polarisation are allowed is related to the axial symmetry of the cloud distribution. 
The switch exists due to a change in the scattering albedo between the soft and the hard bands of the X-ray spectrum. 
Goosmann \& Matt (2011) showed that a more gradual rotation of $\Delta\Psi$ can be obtained in a 
non-relativistic scenario once the distribution of scattering clouds is no longer symmetric. However, 
in this case the rotation can only evolve monotonically with photon energy between two limiting values 
determined by the net soft and hard X-ray polarisation states. Such a scenario could not reproduce the 
S-shape variation of $\Delta\Psi$ seen in the relativistic model.

~\

In conclusion, this work predicts the polarisation degree expected from ``changing look'' AGN in their 
Compton-thin state. The reflection-dominated scenario of NGC~1365 is found to be associated with 
strong polarisation degrees and a smooth variation of polarisation angle across the iron line emission. 
However, if the spectrum of NGC~1365 is dominated by cloud absorption, lower polarisation degrees are expected, 
associated with an orthogonal switch of the polarisation position angle between the soft and the hard 
X-ray band. We strengthen the conclusions drawn in Marin et al. (2012a), despite some fundamental differences 
between MCG-6-30-15 and NGC~1365, and demonstrate that X-ray polarimetry measurement of NGC~1365 could 
discriminate between the two scenario using a hard polarimeter or a large, soft X-ray polarimetric mission.

\section*{Acknowledgments}

We thank the anonymous referee for helpful comments. 
This research was supported by the grants ANR-11-JS56-013-01 in France 
and COST-CZ LD12010 in the Czech Republic.


\bsp

\label{lastpage}

\end{document}